\documentstyle[aps,prb,twocolumn,epsfig]{revtex}
\begin{document}
\twocolumn[\hsize\textwidth\columnwidth\hsize\csname
@twocolumnfalse\endcsname

\title{Observation of a linear temperature dependence of the
critical current density in a Ba$_{0.63}$K$_{0.37}$BiO$_3$ single
crystal$^{\ast}$}
\author{$^a$Hyun-Tak Kim$^{\ast\ast}$, $^a$Kwang-Yong
Kang, $^b$Bong-Jun Kim, $^b$Y. C. Kim, $^c$W. Schmidbauer, and
$^c$J. W. Hodby}
\address{$^a$Telecom. Basic Research Lab., ETRI, Taejon 305-350, Korea\\
$^b$Department of Physics, Pusan National University, Pusan
609-735, Korea\\ $^c$Clarendon Laboratory, University of Oxford,
Parks Road, Oxford OX1 3PU, Great Britain}
\maketitle{}

\begin{abstract}
For a  Ba$_{0.63}$K$_{0.37}$BiO$_3$ single crystal with
$T_c$$\approx$31 K, $H_{c1}{\approx}750~Oe$ at 5 K, and dimensions
3$\times$3$\times$1$~mm^3$, the temperature and field dependences
of magnetic hysteresis loops have been measured within 5-25 K in
magnetic fields up to 6 Tesla. The critical current density is
$J_c(0)\approx$1.5 $\times$ 10$^5 ~A/cm^2$ at zero field and
$1{\times}10^5A/cm^2$ at 1 $kOe$ at 5 K. $J_c$ decreases
exponentially with increasing field up to 10 $kOe$. A linear
temperature dependence of $J_c$ is observed below 25 K, which
differs from the exponential and the power-law temperature
dependences in high-$T_c$ superconductors including the BKBO. The
linear temperature dependence can be regarded as an intrinsic
effect in superconductors.\\ \\
\end{abstract}
]

\narrowtext
 It is well known that Ba$_{1-x}$K$_x$BiO$_3$ (BKBO) with
$T_c{\approx}30$~K is very suitable for research of high-$T_c$
superconductivity, because it has a simple perovskite structure
and characteristics similar to cuprate superconductors. The
superconductivity mechanism and the metal-insulator transition for
BKBO still remain to be clarified. Up to now, although there has
been much research on the superconductivity mechanism for BKBO,
only little of the research was carried out on very high quality
crystals. In this paper, the critical current and its temperature
dependence are investigated by observing magnetic properties of a
high quality BKBO single crystal. The results are compared with
other work published on BKBO and cuprate data characterized by the
power-law and exponential-temperature dependences.

The Ba$_{0.63}$K$_{0.37}$BiO$_3$ single crystal was synthesized by
the electro-chemical method reported elsewhere.$^1$ The size of
the crystal with $T_c\approx$31 K was $3\times3\times1~mm^3$. The
potassium concentration was found to be $x{\approx}0.37$ by
electron-probe microanalysis. The value $H_{c1}{\approx}750~Oe$
was determined at 5 K. The paramagnetic Meissner effect with the
crystal was investigated at low fields.$^2$ The
zero-field-cooled(ZFC), field-cooled(FC) susceptibilities and the
magnetic hysteresis loops were measured by using a magnetometer of
Quantum Design Co.(MPMS7). Before measuring the hysteresis loops,
zero setting for the magnetic field was performed to remove any
remnant field in the superconducting magnet. A magnetic field of 6
Tesla in the c-direction was applied.

Figure 1 shows the ZFC and FC susceptibilities measured at 4 $Oe$
in the virgin-charged superconducting magnet with
 the crystal with $T_c\approx31$ K. The ZFC absolute value evidently exhibits
  no temperature dependence up to 24 K indicated by arrow A and decreases
   rapidly between 24 K and $T_c$. In the case of the ZFC susceptibility, the transition
    width of $T_c$ is ${\triangle}T=7$~K (defined from $T_c$ to 24 K). In the Meissner state,
     the susceptibility is defined as $-4{\pi\chi}_m{\rho}=V/(1-D)$, where
     ${\chi}_m$, $V$ and $D$ are the mass susceptibility, volume fraction and
      demagnetization factor, respectively, while the X-ray density $\rho\approx8~g/cm^3$.
       If $V$ is assumed to be unity and independent of the field orientation,
        $D\approx0.68$ is calculated from the ZFC susceptibility.
        If an ellipsoid of revolution is used to approximate
         the crystal shape with the dimensions mentioned above,
          then $D{\approx}0.65$. This agrees fairly with that calculated
           from the ZFC susceptibility. This agreement indicates that
            the crystal is fully superconducting, nearly single and
             homogeneous.

\begin{figure}
\vspace{1.5cm} \centerline{\epsfxsize=7.4cm \epsfbox{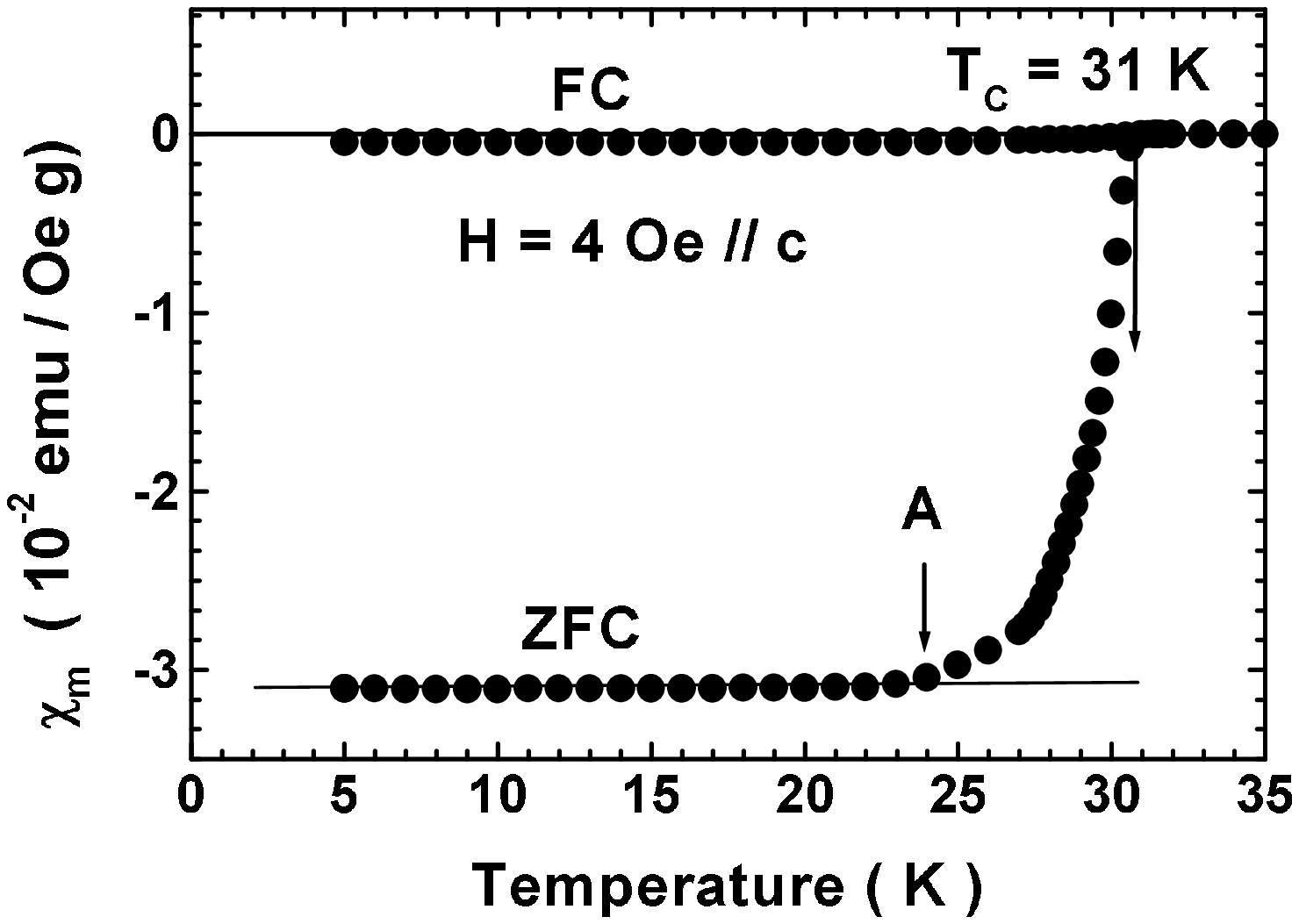}}
\vspace{1.0cm} \caption{The temperature dependence of FC and ZFC
magnetic susceptibilities.}
\end{figure}

Figure 2 (a) and (b) show temperature dependences of a half of the
magnetic hysteresis loops measured up to 6 Tesla in the 5 - 25 K
range; here data are shown only up to 1 Tesla beyond which all
functions are very nearly constant. The loops are typical of those
observed in high $T_c$ superconductors. The field and temperature
dependences of $J_c$ are shown in Figs. 2 (c) and (d). $J_c$ at 5
K at zero field and 1 $kOe$ were evaluated as
$1.5{\times}10^5A/cm^2$ and $1{\times}10^5A/cm^2$, respectively,
by using the Bean critical state model$^3$ applicable to bulk;
$J_c=\frac{10{\triangle}M}{a-\frac{a^2}{3b}}$ in $CGS$ units,
where $a$, $b$ and ${\triangle}M$ are the grain dimensions of a
bulk crystal and the magnetic moment corrected by the
demagnetization factor, respectively. The magnetic field
dependence of $J_c$ decreases exponentially with increasing field
up to 10 $kOe$. The temperature dependence of $J_c$, as shown in
Fig. 2 (d), is linear below $\sim$25 K indicated as arrow A in
Fig. 1. This indicates that the superconducting condensed state
below $T_c$ has a linear temperature dependence for $J_c$. The
linear dependence differs from the published results which follow
a power-law$^{4,5}$ for the BKBO and an exponental dependence$^6$
for a Hg system. The cause of the difference is that the absolute
values of ZFC susceptibilities at low temperatures in these papers
are not constant but decrease with increasing temperature$^{4,6}$;
$i.e.$, this indicates that the nonlinearity (or power law) may be
attributed to a pinning effect due to impurities in the crystals.
In addition to the above crystal, the linearity for another high
quality crystal was observed by the same experimental method.

Finally, we suggest that the linear temperature dependence of
$J_c$ is an intrinsic effect in superconductors.

\begin{figure}
\vspace{1.0cm}\centerline{\epsfxsize=7.0cm\epsfbox{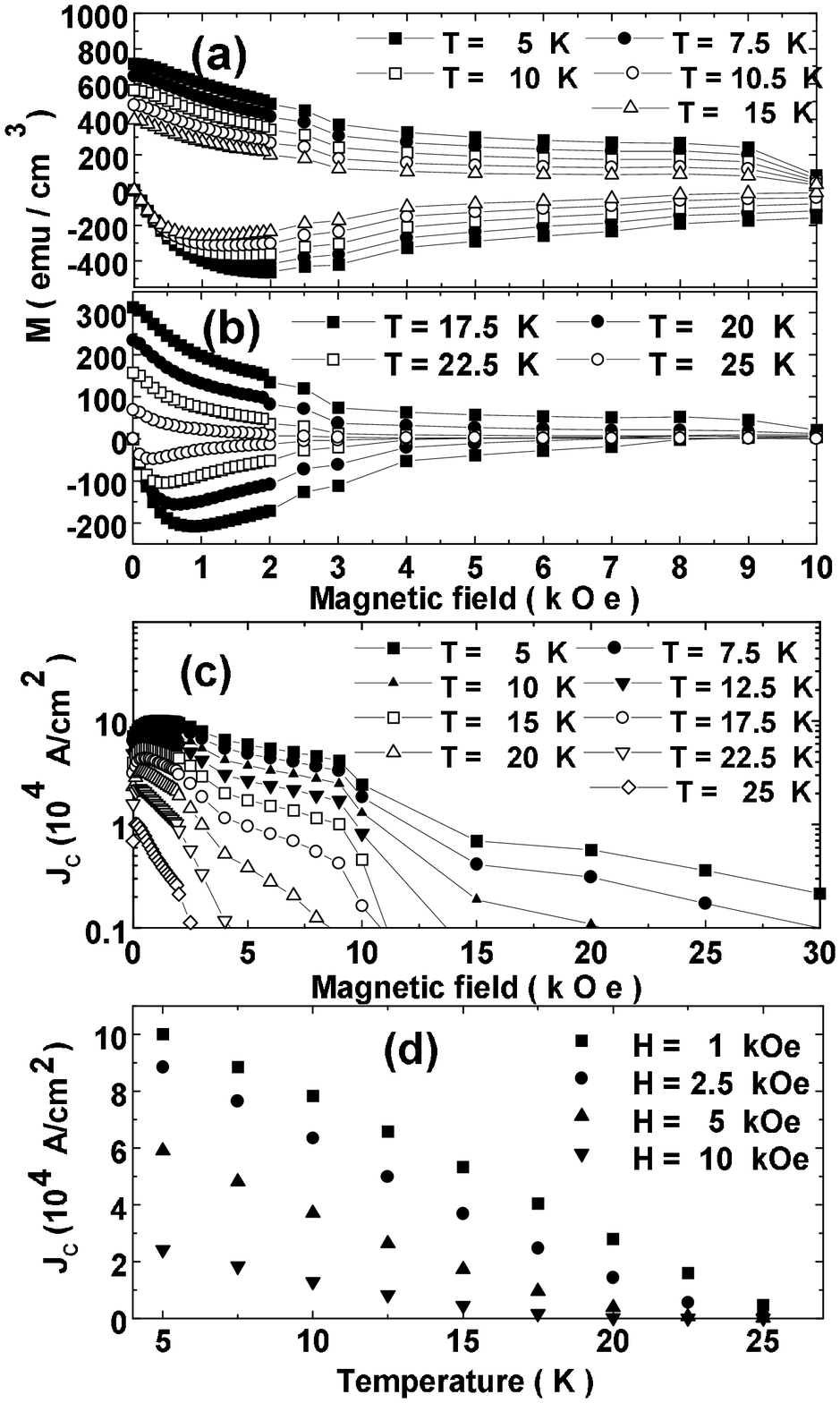}}
\vspace{1.0cm}\caption{(a) and (b) the temperature dependence of a
half of magnetic hysteresis loops; (c) the field dependence of
$J_c$; (d) the temperature dependence of $J_c$.}
\end{figure}

\end{document}